\documentclass{aa}  
\usepackage{graphicx}
\bibpunct{(}{)}{;}{a}{}{,}
\usepackage[varg]{txfonts}
\usepackage{natbib,twoopt}
\usepackage[english]{babel}
\usepackage{float}
\usepackage{physics}
\usepackage{placeins}
\usepackage{txfonts}
\usepackage{xcolor}%
\usepackage{textcomp}%
\usepackage[colorlinks=true,citecolor=blue]{hyperref} 
\bibpunct{(}{)}{;}{a}{}{,} 
\makeatletter
\newcommandtwoopt{\citeads}[3][][]{\href{http://adsabs.harvard.edu/abs/#3}
{\def\hyper@linkstart##1##2{}
\let\hyper@linkend\@empty\citealp[#1][#2]{#3}}}
\newcommandtwoopt{\citepads}[3][][]{\href{http://adsabs.harvard.edu/abs/#3}
{\def\hyper@linkstart##1##2{}
\let\hyper@linkend\@empty\citep[#1][#2]{#3}}}
\newcommandtwoopt{\citetads}[3][][]{\href{http://adsabs.harvard.edu/abs/#3}
{\def\hyper@linkstart##1##2{}
\let\hyper@linkend\@empty\citet[#1][#2]{#3}}}
\newcommandtwoopt{\citeyearads}[3][][]
{\href{http://adsabs.harvard.edu/abs/#3}
{\def\hyper@linkstart##1##2{}
\let\hyper@linkend\@empty\citeyear[#1][#2]{#3}}}
\makeatother
\bibpunct{(}{)}{;}{a}{}{,}
\usepackage{booktabs}
\usepackage{multirow}
\usepackage{orcidlink}
\usepackage{graphicx}

\usepackage{txfonts}

\newcommand{\giano}{GIANO-B\xspace}
\newcommand{\sysrem}{\textsc{sysrem}\xspace}

\begin{document} 

\title{Cross-correlation transmission spectroscopy of ultra-hot Jupiters WASP-189b, HAT-P-57b, KELT-17b, and KELT-21b with GIANO-B}

    \titlerunning{Atmospheres of four UHJs}
   
   \author{P. Meni-Gallardo \inst{\ref{ins:iac} ,\ref{ins:ull}}\orcidlink{0009-0001-7943-0075}
   \and
   J. Orell-Miquel \inst{\ref{ins:texas},\ref{ins:iac} ,\ref{ins:ull}} \orcidlink{0000-0003-2066-8959}
   \and Gareb Fernández-Rodríguez \inst{\ref{ins:iac},\ref{ins:ull}} \orcidlink{0000-0003-0597-7809}
   \and
   H. Parviainen\inst{\ref{ins:iac},\ref{ins:ull}} \orcidlink{0000-0001-5519-1391}
   \and 
   M. Basilicata\inst{\ref{ins:oato}}
   \and 
   E. Pallé \inst{\ref{ins:iac},\ref{ins:ull}} \orcidlink{0000-0003-0987-1593}   \and 
   M. Stangret\inst{\ref{ins:oapd}} \orcidlink{0000-0002-1812-8024}  
   \and 
   I. Carleo\inst{\ref{ins:oato}} \orcidlink{0000-0002-0810-3747}
   \and 
   P. Giacobbe\inst{\ref{ins:oato}} \orcidlink{0000-0001-7034-7024}
   }

\institute{
        \label{ins:iac}Instituto de Astrof\'isica de Canarias (IAC), 38205 La Laguna, Tenerife, Spain
        \and
        \label{ins:ull}Departamento de Astrof\'isica, Universidad de La Laguna (ULL), 38206 La Laguna, Tenerife, Spain
        \and
        \label{ins:texas}Department of Astronomy, University of Texas at Austin, 2515 Speedway, Austin, TX 78712, USA
        \and
        \label{ins:oato}INAF -- Osservatorio Astrofisico di Torino, Via Osservatorio 20, 10025 Pino Torinese, Italy
        \and
        \label{ins:oapd}INAF -- Osservatorio Astronomico di Padova, Vicolo dell'Osservatorio 5, 35122, Padova, Italy
        }

   \date{Received DD Month 2025 / Accepted DD Month YYYY}

\abstract{Since the discovery of the first exoplanet, significant efforts have been made to characterise their atmospheres. Ultra-hot Jupiters (UHJs) are of particular interest due to their extended and hot atmospheres. Although previous studies have focused on the detection of atomic species at optical wavelengths, near-infrared (NIR) observations offer the potential to detect molecules. In our study, we applied the cross-correlation technique to NIR transmission spectra from \giano. The analysis focuses on the search for H$_2$O, CO, CO$_2$, CH$_4$, HCN, and FeH molecular signals in the atmospheres of four UHJs: HAT-P-57 b, KELT-17 b, KELT-21 b, and WASP-189 b. For the first time, we report results on the NIR transmission spectra of KELT-17b, KELT-21b, and WASP-189b. We report a tentative detection ($3.8\sigma$) of H$_2$O in HAT-P-57 b and a detection ($5.3\sigma$) of FeH in KELT-17~b, which is the third FeH detection ever in a UHJ and with the lowest equilibrium temperature. No molecular signals were found in KELT-21b and WASP-189b, or for other molecules in HAT-P-57b and KELT-17b.

The cross-correlation results for HAT-P-57 b, KELT-17 b, KELT-21 b, and WASP-189 b in transmission align with the species detected in the UHJ population. This work underscores the need for further observations to confirm and expand the transmission study of UHJs in the NIR, and the capabilities of high-resolution spectrographs on 4-m-class telescopes.

}
\maketitle
\nolinenumbers

\section{Introduction}

The advent of high-resolution spectrographs has enabled the detection and characterisation of exoplanetary atmospheres by leveraging the differential radial velocities of the planet, host star, and Earth. Among the various exoplanet types, ultra-hot Jupiters (UHJs) and hot Jupiters (HJs) are the most accessible targets for ground-based atmospheric studies. High temperatures, short orbital periods, and extended atmospheres make them ideal for transmission spectroscopy. UHJs, defined as gas giants with equilibrium temperatures exceeding 2000~K \citep{2018Parmentier}, are the focus of this study. Their extended atmospheres and high infrared brightness make them well suited for high-resolution spectroscopy. Due to their extreme temperatures, clouds are unlikely to form on their daysides \citep{2019Helling, 2021Helling, 2023Helling}, except for highly refractory species such as aluminium and titanium \citep{2017Wakeford}. By analysing the atmospheric chemistry of UHJs, we can constrain their formation scenarios and evolutionary pathways \citep{2021Turrini}.

One major challenge in UHJ studies is the potential dissociation of molecular species due to high temperatures \citep{2018Parmentier, 2020Mansfield}. Consequently, most atmospheric detections have been made in the optical range \citep{Stangret2022, Stangret2020, Stangret2021, Casasayas-Barris2019}, where species such as Fe and H$^-$ dominate the opacity \citep{2020JensHoe, 2024Deibert, 2025Vaulato}. High-resolution optical spectroscopy surveys have also detected numerous atomic and ionic species \citep{2021merrit, prinoth2022titanium, pelletier2023vanadium}.

However, with recent advances in instrumentation and analysis techniques, molecular detections in the optical and near-infrared (NIR) have become more common. Several studies have reported molecular detections in UHJs and HJs using ground-based telescopes, such as H$_2$O in  MASCARA-2~b/KELT-20~b \citep{2022Fu}, WASP-33~b \citep{2024Yang}, and HD~189733~b \citep{Birkby2013a}; CO in WASP-33~b \citep{yan2022detection}, HD~189733~b \citep{cabot2019robustness, blain2024formally}, and WASP-189~b \citep{yan2022detection}; CO$_2$ in WASP-69$~$b \citep{2022Guilluy}; CH$_4$ in 2M0437$~$b \citep{2023Gaidos}; HCN in HD$~$209458$~$b \citep{giacobbe2021five} and WASP-80~b \citep{2022Ilaria}; FeH in Luhman$~$16$~$A\citep{2025Regt}; TiO in WASP-19$~$b \citep{2021Seda}; VO in TOI-1518$~$b \citep{2025simonin}; and OH in WASP-18$~$b \citep{2023brogi}.

In this context, the focus of this work is the search for molecular species in the atmospheres of four UHJs: HAT-P-57~b, KELT-17~b, KELT-21~b, and WASP-189~b. Their stellar and planetary parameters are summarised in Table~\ref{tab:parameters}. Their optical transmission spectra were previously analysed by \citet{Stangret2021}. Here, we extend their study into the NIR, allowing for a direct comparison between different wavelength regimes.

\begin{table*}[hpbt]
    \caption{Stellar and planetary parameters.}
    \centering
    \begin{tabular*}{\linewidth}{@{\extracolsep{\fill}} lcccc}
        \toprule
        \toprule
         & HAT-P-57 & KELT-17 & WASP-189 & KELT-21 \\
        \multicolumn{5}{l}{\emph{Equatorial coordinates}} \\
        \midrule
        RA (J2000) & 18:18:58.4 & 08:22:28.0 & 15:02:44.9 & 20:19:12.0 \\
        Dec (J2000) & +10:35:50.1 & +13:44:07.0 & -03:01:52.9 & +32:34:52.0 \\
        \\
        \multicolumn{5}{l}{\emph{Stellar parameters}} \\
        \midrule
        Spectral type & A8V & A & A6IV-V & A8V \\[2pt]
        $M_* [M_\odot]$ & $1.47 \pm 0.12$ & $1.635 \pm 0.060$ & $2.030 \pm 0.066$ & $1.458 \pm 0.028$ \\
        $R_* [R_\odot]$ & $1.50 \pm 0.05$ & $1.645 \pm 0.055$ & $2.36 \pm 0.03$ & $1.638 \pm 0.034$ \\
        $K_\mathrm{s}$ [m/s] & $<215.2$ & $131 \pm 34$ & $182 \pm 13$ & $<399.6$ \\
        $\log g$ [dex] & $4.251 \pm 0.018$ & $4.220 \pm 0.023$ & $3.9 \pm 0.2$ & $4.173 \pm 0.014$ \\
        $T_\mathrm{eff} [K]$ & $7500 \pm 250$ & $7454 \pm 49$ & $7996 \pm 99$ & $7598 \pm 8$ \\
        \\
        \multicolumn{5}{l}{\emph{Planetary parameters}} \\
        \midrule
        Orbital period [d] & $2.46529488$ & $3.08017985$ & $2.7240308$ & $3.61276958$ \\
        $T_0$ [d] & $2457598.49926$ & $2459440.791304$ & $2456706.4566$ & $2458881.93965$ \\
        $M_\mathrm{pl} \,[M_\mathrm{J}]$ & $1.41 \pm 1.52$ & $1.31 \pm 0.28$ & $1.99 \pm 0.15$ & $<3.91$ \\
        $R_\mathrm{pl}\, [R_\mathrm{J}]$ & $1.413 \pm 0.054$ & $1.525 \pm 0.062$ & $1.619 \pm 0.021$ & $1.586 \pm 0.039$ \\
        $i \,[^\circ]$ & $88.26 \pm 0.85$ & $84.87 \pm 0.44$ & $84.03 \pm 0.14$ & $86.46 \pm 0.36$ \\
        $V_\mathrm{sys}$ [km/s] & $-9.6$ & $28.0$ & $-24.465$ & $-13.0$ \\
        $T_{14}$ [h] & $3.499$ & $3.475$ & $4.3336$ & $4.105$ \\
        $a$ [AU] & $0.0406 \pm 0.0011$ & $0.04881 \pm 0.00063$ & $0.05053 \pm 0.00098$ & $0.05224 \pm 0.00034$ \\
        $K_\mathrm{p}$ [km/s] & $180 \pm 5$ & $171 \pm 37$ & $195 \pm 21$ & $156 \pm 72$ \\
        $T_\mathrm{eq}$ [K] & $2200 \pm 76$ & $2087 \pm 32$ & $2641 \pm 34$ & $2051 \pm 29$ \\
    \bottomrule
    \bottomrule
    \end{tabular*}
    \tablefoot{The references for each system are as follows: 1) HAT-P-57/HAT-P-57~b: OP and $\rm T_0$ from \cite{kokori2023exoclock}; RA, Dec, $\rm M_*$, $\rm R_*$, $\rm K_s$, $\rm log_g$, $\rm T_{eff}$, $\rm R_{pl}$, $\rm i$, $\rm T_{14}$, $\rm a$, and $\rm T_{eq}$ from \cite{hartman2015hat}; $\rm M_{pl}$ from \cite{stassun2017accurate}; $K_P$ from \cite{2024Mopys}; $\rm V_{sys}$ from \citet{prusti2016gaia}, \citet{brown2018gaia}, \citet{cropper2018gaia}, and \citet{katz2019gaia};
    2) KELT-17/KELT-17~b:  OP and $\rm T_0$ from \cite{kokori2023exoclock};  RA, Dec, $\rm M_*$, $\rm R_*$, $\rm K_s$, $\rm log_g$, $\rm T_{eff}$, $\rm R_{pl}$, $\rm i$, $\rm T_{14}$, $\rm a$, and $\rm T_{eq}$, $\rm M_{pl}$ and $\rm V_{sys}$ from \cite{zhou2016kelt}; $K_P$ derived from \cite{stassun2017accurate};
    3) KELT-21/KELT-21~b: OP and $\rm T_0$ from \cite{kokori2023exoclock}; RA, Dec, $\rm M_*$, $\rm R_*$, $\rm K_s$, $\rm log_g$, $\rm T_{eff}$, $\rm R_{pl}$, $\rm i$, $\rm T_{14}$, $\rm a$, and $\rm T_{eq}$, $\rm M_{pl}$ , and $\rm V_{sys}$ from \cite{johnson2018kelt}; $K_P$ derived from \cite{stassun2017accurate}; and
    4)  WASP-189/WASP-189~b: OP from \cite{ivshina2022tess};$\rm T_0$ from \cite{kokori2023exoclock};  RA, Dec, $\rm M_*$, $\rm R_*$, $\rm K_s$, $\rm log_g$, $\rm T_{eff}$, $\rm R_{pl}$, $\rm i$, $\rm T_{14}$, $\rm a$, and $\rm T_{eq}$ from \cite{anderson2018wasp189bultrahotjupitertransiting}; $K_P$ derived from \cite{stassun2017accurate}.}
    \label{tab:parameters}
\end{table*}

\section{Observations}
\label{sec: Obs}

\begin{figure*}[!hpbt]
    \centering
    \includegraphics[width=0.95\columnwidth]{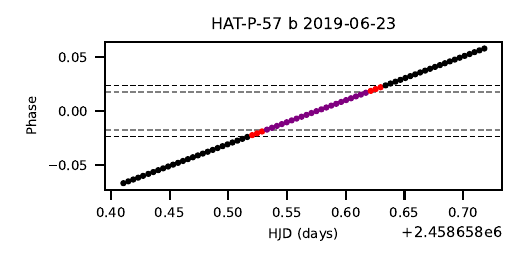}
    \includegraphics[width=.95\columnwidth]{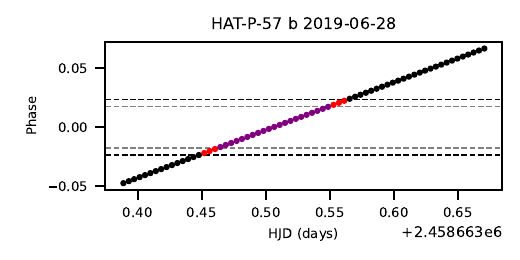}\\
    \includegraphics[width=.95\columnwidth]{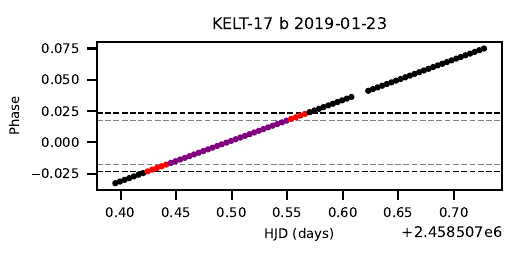}
    \includegraphics[width=.95\columnwidth]{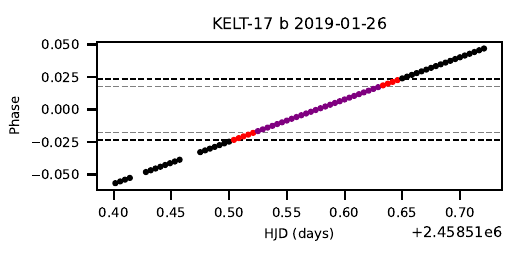}\\
    \includegraphics[width=.95\columnwidth]{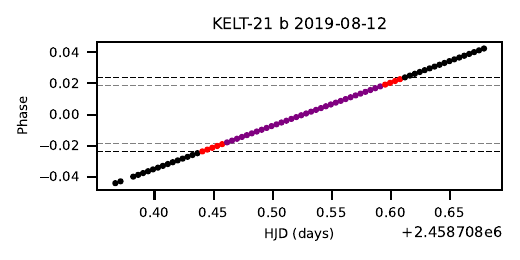}
    \includegraphics[width=.95\columnwidth]{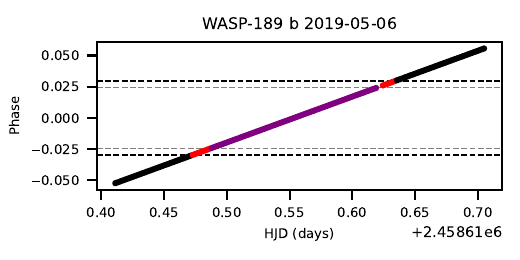}
    \caption{Orbital phase of  each planet. Each dot represents a single spectral frame, so this diagram shows the evolution of the planet's position over time. The dashed black lines represent the moments when the planets began and ended the transit, $\rm T_1$ and $\rm T_4$. The dashed grey lines show the moments when the planet's surface is completely inside and outside the star's surface, $\rm T_2$ and $\rm T_3$. The black points are the out-of-transit frames, the red ones are the transit's beginning and end, and the purple dots represent the in-transit frames. The top images are for HAT-P-57 b on 23 and 28 June 2019, the middle images are for KELT-17 b on 23 and 26 January 2019, and the bottom images are for KELT-21 b and WASP-189 b in that order.}
    \label{fig:phaseall}
\end{figure*}

We analysed observations of UHJs taken with the \giano-B spectrograph \citep{oliva_giano_2012} mounted at the 3.6\,m Telescopio Nazionale Galileo (TNG) at Observatorio Roque de los Muchachos (ORM) in La Palma, Spain. The observations were done using the standard ABAB nodding configuration \citep{2016claudi}.
Table~\ref{tab:observations} provides the observing log and the phase diagram related to each night can be seen in Fig. \ref{fig:phaseall}. Below, we provide the key details about the planetary systems and their observations.

\begin{itemize}
    \item HAT-P-57~b \citep{hartman2015hat} has an equilibrium temperature of 2200 K, a radius of $\mathrm{1.4}\,R_J$, and a mass of $\mathrm{1.4}\,M_J$, and it orbits an A8\,V main sequence star with a period of 2.5 days. We observed it on two nights (CAT19A\_97, PI: Casasayas-Barris), 23 and 28 June 2019, and independently analysed the results of each night. For 23 June 2019 data (night 1), we collected 74 spectral frames with exposure times of 300\,s with a mean signal-to-noise ratio (S/N) of 19. For 28 June 2019 data (night 2), we collected 68 frames with exposure times of 300\,s with a mean S/N of 23.
    \item  KELT-17~b \citep{zhou2016kelt} has an equilibrium temperature of 2087 K, a radius of $\mathrm{1.3}\,R_J,$ and a mass of $\mathrm{1.5}\,M_J$, and it orbits a fast-rotating A star with a period of 3.08 days. We observed it on two nights (CAT18B\_62, PI: Casasayas-Barris). We obtained 81 frames of 300\,s exposure times with a mean S/N of 30 on night 1 (23 January 2019), and 72 spectral frames of 300\,s exposure times with a mean S/N of 29 on night 2 (26 January 2019).
    \item KELT-21~b \citep{johnson2018kelt} has an equilibrium temperature of 2051 K, a radius of $\mathrm{1.5}\,R_J,$ and an upper mass of $< \mathrm{3.9}\,M_J$, and it orbits an A8\,V star with a period of 3.61 days. We obtained one transit (CAT19A\_97, PI: Casasayas-Barris) on the night of 12 August 2019, taking 76 spectral frames of 300\,s exposure times with a mean S/N of 23.
    \item WASP-189~b \citep{anderson2018wasp189bultrahotjupitertransiting} is the hottest target of this work with an equilibrium temperature of 2614$\,$K. The planet has a radius of $\mathrm{1.6}\,R_J$ and a mass of $\mathrm{2}\,M_J$, and orbits a rapidly rotating A6\,IV-V star with a period of 2.72 days. We observed one transit on 6 May 2019 (CAT19A\_97, PI: Casasayas-Barris) obtaining 156 spectral frames of 100\,s exposure times with a mean S/N of 66.

\end{itemize}

\begin{table*}[hpbt]
    \caption{Summary of the observations.}
    \centering
    \begin{tabular*}{\linewidth}{@{\extracolsep{\fill}} llllllll}
    \toprule \toprule
        Object & Night & Date of observation & $\rm T_{exp} [s] $ & $\rm N_{in}/N_{obs}$  &  Mean $\rm S/N$ & Used orders \\ [2pt]
        \midrule
        HAT-P-57~b & 1 & 2019-06-23 & 300 & 21/74 & 19 & 17/50\\
        HAT-P-57~b & 2 & 2019-06-28 & 300 &21/68 & 23  & 17/50\\
        KELT-17~b & 1 & 2019-01-23 & 300 & 26/78 & 30 & 26/50\\
        KELT-17~b & 2 & 2019-01-26 & 300 & 26/72 & 29 & 31/50 \\
        KELT-21~b & 1 & 2019-08-12 & 300 & 32/74 & 23 & 16/50\\
        WASP-189~b & 1 & 2019-05-06 & 100 & 71/158 & 66 & 24/50\\
    \bottomrule \bottomrule
    \end{tabular*}
    \label{tab:observations}
\end{table*}

\section{Methods}
\label{sec: Methods}

\subsection{Atmosphere modelling} 
\label{ssec:atm}

We computed synthetic atmospheric models using the \texttt{petitRADTRANS} code \citep{Molli_re_2019} to simulate the expected molecular signatures of the exoplanets' atmospheres. The models were only computed for the molecules of H$_2$O \citep{H2O_Poka}, CO \citep{CO_Li}, CO$_2$ \citep{CO2}, CH$_4$ \citep{CH4}, HCN \citep{HCN}, FeH \citep{BERNATH2020106687}, TiO \citep{2021TiO} based on the original \cite{mckemmish2019exomol}, VO \citep{Molli_re_2019} from the original list from \cite{mckemmish2016exomol}, and OH \citep{2018OH} as they are the most common species that exhibit strong absorption features within \giano-B's spectral range. For simplicity, we used uniform solar elemental abundance \citep{Asplund} profiles. Since the atmospheric abundance profiles are likely to be non-uniform with altitude, this can give biased results when retrieved using models assuming a well-mixed atmosphere \citep{Lesjak2024, 2025sanchez}. For each molecular species, we manually estimated the abundance based on a solar composition, with ranges between $10^{-4}$ to $10^{-10}$ depending on the molecule. This enhancement was made to improve the S/N of the cross-correlation (CC).

\begin{figure*}[hpbt]
    \centering
    \includegraphics[width=0.95\linewidth]{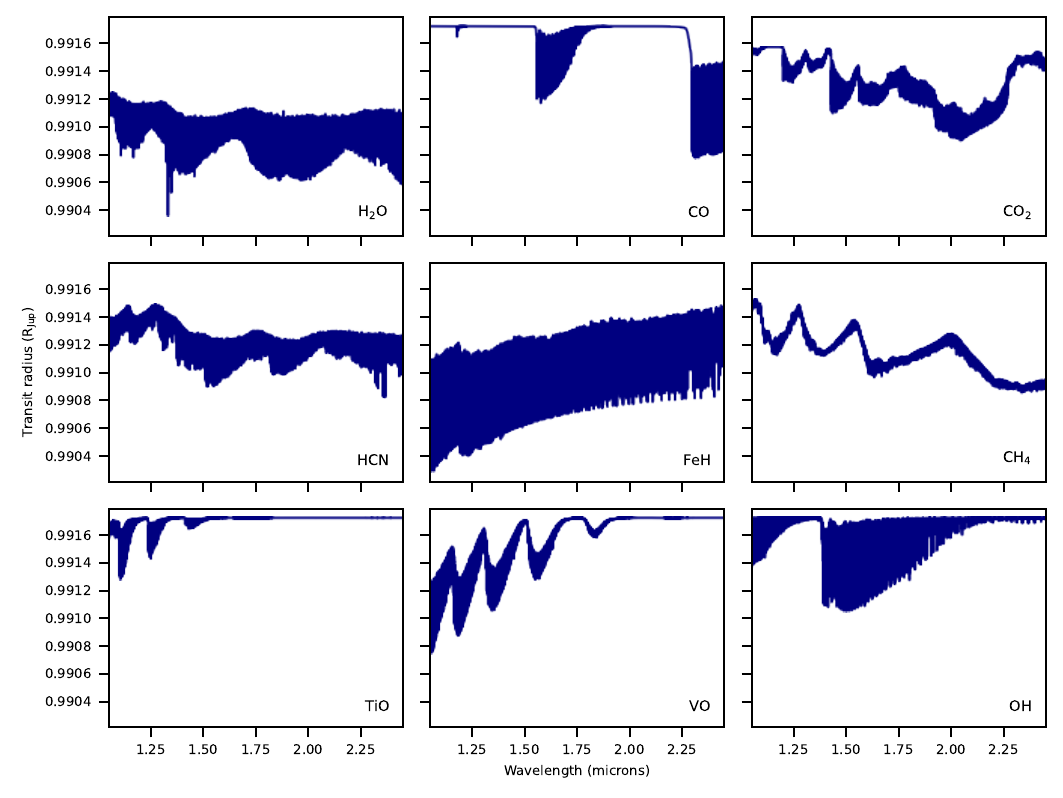}
    \caption{Atmospheric models for HAT-P-57~b. Left from top to bottom: Spectral features of H$_2$O, CO, and CO$_2$. Middle from top to bottom: Spectral signatures of CH$_4$, HCN, and FeH. Right from top to bottom: Spectral traces of TiO, VO, and OH, generated with \texttt{petitRADTRANS}.}
    \label{fig:123}
\end{figure*}

The equilibrium temperature of each planet was used as an input parameter, assuming an isothermal atmosphere and a reference pressure of 0.01 bar. Figure~\ref{fig:123} shows all six molecular models for HAT-P-57$~$b for illustrative purposes. The models were initially generated at a high spectral resolution (R$\sim$1\,000\,000) and subsequently downsampled to match the resolving power of \giano-B (R$\sim$50\,000) to ensure consistency with observational data. This downsampling was performed following the same methodology as used in \cite{Stangret2021}. The method is based on the interpolation between the wavelength provided by the instrument, after useless orders are removed, and the computed molecular model. The orders considered as problematic were those that show problems in the shifting process during nodding or those that are aligned with telluric bands.

\subsection{Data reduction}

Our data consist of spectroscopic time series stored as three-dimensional arrays and auxiliary information such as the orbital phase of each exposure. The initial reduction was performed with the offline version of \giano online flux information organizer (GOFIO) pipeline \citep{rainer_introducing_2018}, designed for the \giano spectrograph. GOFIO optimally extracts the spectral trace from the two nodding positions (A and B) and provides a preliminary wavelength calibration based on U-Ne lamp exposures. However, this initial wavelength calibration does not achieve the accuracy required for CC analyses \citep{giacobbe2021five, 2022Ilaria}, so we refined the wavelength solution using a set of Python scripts used in \citet{giacobbe2021five}. The code workflow consists of four main steps: 

\begin{enumerate} 
    \item Data extraction and organisation: The first routine extracts spectral data and header information from the fits files reduced by GOFIO (e.g. air mass and S/N), organises the spectra into a cube (50 orders $\times$ the number of frames for each dataset $\times$ 2048 pixels), and outputs the initial wavelength solution as a $50 \times 2048$ matrix. 
    \item Orbital phase and barycentric correction: The second routine calculates the orbital phase for each frame and computes (but does not apply) the barycentric correction using the system parameters (transit epoch, period, systemic velocity, transit boundaries, and radial velocity semi-amplitude). This correction is essential for accurately aligning the spectral data during the CC analysis. 
    \item Spectral alignment: In the third step, the routine aligns the spectra within each order with the mean spectrum of that order. This alignment corrects for any drift on the \giano detector. The aligned data cube is saved for further analysis and diagnostic plots are generated to visualise the drift. 
    \item Refined wavelength calibration: Finally, a more precise wavelength calibration is performed using a template of telluric lines, generated via the European Southern Observatory (ESO) Sky Model Calculator \citep{noll2012atmospheric, jones2013advanced}, producing an improved wavelength solution to compute the cross-correlation functions (CCFs). 
\end{enumerate}

\subsection{Cross-correlation analysis}

Cross-correlation analysis begins with the removal of bad spectral orders identified from the pre-processing pipeline diagnostic plots. Some orders are discarded because they are severely affected by telluric absorption, while others are excluded due to instrumental and observational artefacts or processing issues.
Additionally, we mitigate the effects of outliers (e.g. hot and cold pixels) by applying a median filter instead of a polynomial fit \citep{Stangret2021}, which can be sensitive to extreme values in flux. Specifically, we subtract a running median from each spectrum and flag pixels that deviate by more than $5\sigma$ in spectral dimension from the residual distribution. In addition, pixels with values below one are classified as cold pixels to prevent instabilities during normalisation.

Each spectral order of each spectrum is then normalised by splitting it into 50 segments. In each segment, the 50 highest flux values are used to fit a polynomial as a function of wavelength. The division by this polynomial removes long-term trends, which results in a normalised spectrum.

To further remove systematic effects -- primarily telluric contamination and residual stellar signals --, we apply the \sysrem algorithm \citep{tamuz_correcting_2005}. Our implementation, based on previous studies \citep{Birkby2013a, Birkby2017a, Stangret2020, Stangret2021, Stangret2022}, typically requires eight to nine iterations to remove the telluric and stellar contribution but not any possible planetary signal.

Our synthetic model spectra have a much higher resolution than the data and are not segmented into orders, thus we binned and normalised them using the same procedure applied to the data. This method ensures consistency in the CC analysis.

\begin{figure*}[!hpbt]
    \centering
    \includegraphics[width=.93\linewidth]{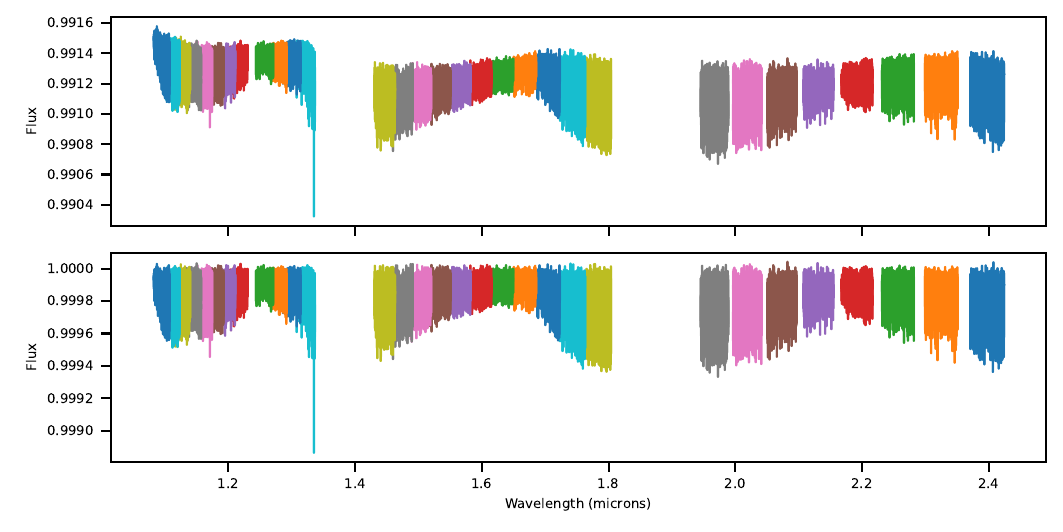}
    \caption{$\rm H_2O$ model created with petitRADTRANS cut in orders to make it fit with our data on the top and the same model normalised on the bottom.}
    \label{fig:modhat}
\end{figure*}

We defined a radial velocity grid from $-225$ to $225\,\mathrm{km\,s^{-1}}$, in steps of the resolution element ($2.7\,\mathrm{km\,s^{-1}}$ \citealp{brogi_exoplanet_2018}). We then performed the CC using the \texttt{crosscorrRV} module from the pyAstronomy package \citep{pya}, and modelled the resulting signal with a polynomial fit. The analysis outputs include a residual map on the Earth rest frame that is smoothed by fitting a third-order polynomial to each CC, a semi-amplitude velocity ($K_\mathrm{p}$) map (that displays the CC signal as a function of the planetary semi-amplitude and systemic velocity in the planet rest frame), and a corresponding CCF map for the pertinent $K_\mathrm{p}$. We selected the system iteration in which the telluric lines were visually significantly removed from the spectroscopic time series for each night. The same \sysrem iteration was used for all spectral orders and all molecules.

\subsection{Injection and recovery}
We tested the detectability of the different species using injection and recovery tests, obtaining $3\sigma$ sensibility curves for the mass-fraction abundances (abundances hereafter) on the basis of solar-abundance models (see Figure \ref{fig:context2}). The artificial atmospheric signal was injected at $-K_p$ with an offset of $\Delta v=-50\,\mathrm{km/s}$, where no atmospheric features are expected. This approach was extensively used in these tests and ensures no contamination of other sources (\citealp{Cont2021}). A model template was created and multiplied by the in-transit normalised spectra. The resulting matrix were then analysed with the same procedure as the data.

We used the same planet parameters and \sysrem iterations as in the CC analysis for consistency in the injection of each molecule. We then varied the abundances to map the sensitivity of the data. The injected signal only includes the molecule being studied and filling gases, namely H2 and He. The filling gases have varying abundances but maintain a fixed He/H2 ratio, depending on the injected mass mixing ratio. No other molecules are injected simultaneously.

\section{Results and the injection and recovery test}
\label{sec: Results}

In this section, we present the results from the CCF analyses of the \giano-B datasets. We also present the sensitivity of our data with injection and recovery tests (IRTs). We took into account the sensitivities of the IRTs (see Figure \ref{fig:context2}) to interpret the CCF results for each night.

\subsection{HAT-P-57~b}

\begin{figure}[!hpbt]
    \centering
    \includegraphics[width=\linewidth]{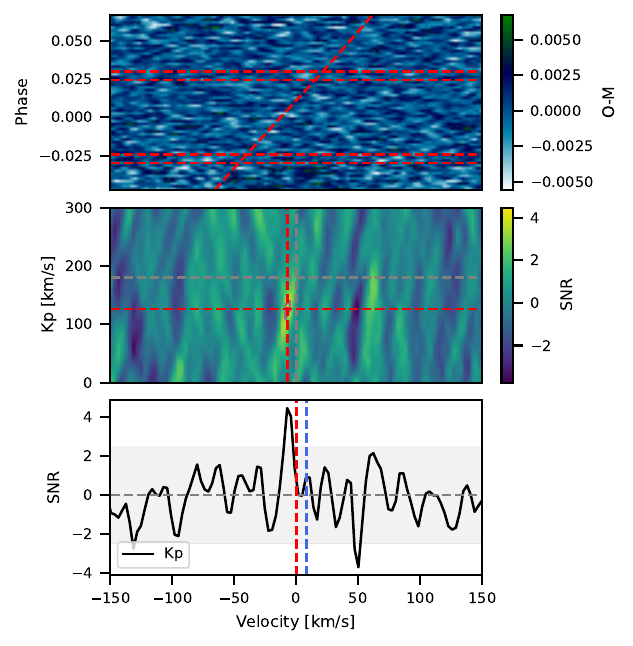}
    \caption{Results of the HAT-P-57 b on the night of 23 June 2019 CC analysis for H$_2$O. Top: Residual map obtained from the cross-correlation. The tilted dashed red line represents the molecule fingerprint evolution during the observation, and the horizontal dashed red lines represent the different phases of the transit; the first line is $\rm T_1$, so the next scales to $\rm T_2$. 'O-M' means Observation minus Model . If tellurics appeared, they would be visible at rv = 0 km/s as these residuals are in the Earth rest frame. Middle: $\rm K_P$ map showing the cross-correlated signal distribution in terms of the semi-amplitude velocity, the systematic velocity, and the S/N. The vertical dashed lines are centred on zero due to the reference frame being on the planet, and the horizontal line marks the planet's $\rm K_P$, the dashed red lines represent the peak of the detection. Bottom: 'Slice' of the CCF that belongs to the $\rm K_P$ where the planet has the highest S/N. The black line is the signal detected after applying cross-correlation, the red line marks the 0 RV, the dashed blue line marks where the tellurics should appear, and the grey contour is limited to $\pm 2.5$ S/N, which is our initial detection criterion.} 
    \label{fig:H2Ohat2}
\end{figure}

For night 1, we used 17 of the 50 available orders and selected the fifth iteration of \sysrem. The CC analysis revealed a detection for H$_2$O (Fig.~\ref{fig:H2Ohat2}) at a $3.8\sigma$ significance level, but not at the expected $K_\mathrm{p}$ position. This offset is not necessarily inconsistent with the planetary origin of the signal due to the high uncertainty of the planet's semi-amplitude velocity. No other molecular signatures ($\rm CO,\, CO_2,\, HCN,\,FeH,\, CH_4,\,TiO,\,VO,\,and\,OH$) were detected.

\begin{figure}[!hpbt]
    \centering
    \includegraphics[width=\linewidth]{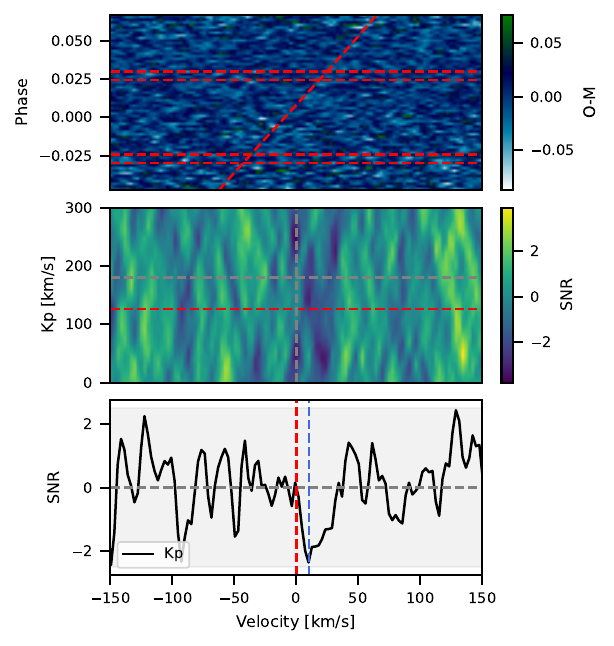}
    \caption{Result of the CC  analysis for $\rm H_2O$ on HAT-P-57 on night of 28 June 2019. The interpretation of the dashed lines is as in Fig. 4.}
    \label{fig:CO2hat1}
\end{figure}

\begin{figure}[!hpbt]
    \centering
    \includegraphics[width=\linewidth]{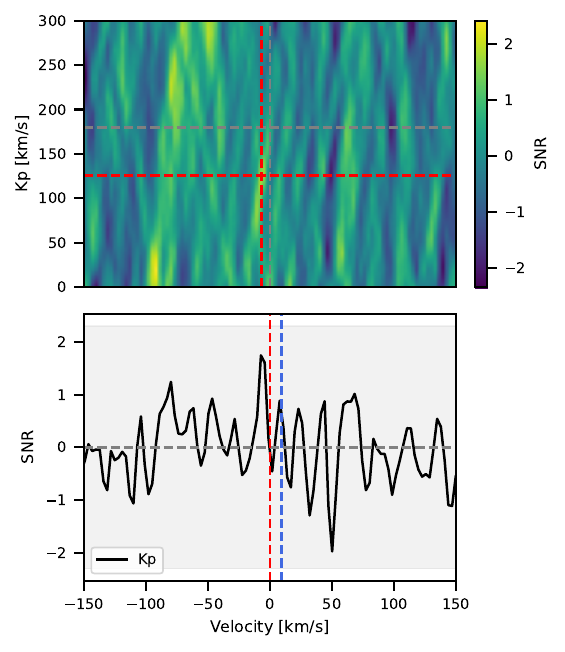}
    \caption{Result of the combination HAT-P-57~b's nights CC  analysis for $\rm H_2O$. The interpretation of the dashed lines is as in Fig. 4.}
    \label{fig:CO2hatcomb}
\end{figure}

For night 2,  17 spectral orders were also used and the highest S/N was achieved in the sixth iteration of \sysrem. We did not recover the H$_2$O signal on night 2 (Fig~\ref{fig:CO2hat1}), as well as the other inspected molecules; this non-detection could be produced by the tellurics or systematics during the second night. If we combine both nights, the H$_2$O signal is not detected with significance (see Figure \ref{fig:CO2hatcomb}) and the other molecules remain undetected. Moreover, the large uncertainty in the planetary mass propagates into the semi-amplitude velocity ($K_\mathrm{p} = 180 \pm 5\,\mathrm{km,s^{-1}}$) and challenges our ability to constrain the planet's radial velocity signal in the CC maps and to interpret the H$_2$O tentative signal.

The IRTs indicate that night 1 is more sensitive to the detection of $\rm H_2O$ than night 2. Although the different quality of the data may explain the discrepancy between the nights, we cautiously claim the $\rm H_2O$ signal as a tentative detection despite its significance of $3.8\sigma$ on the first night.
FeH, TiO, VO, and OH should be detectable on both nights for abundances greater than $\sim$$10^{-7}$ according to the IRTs, but the signal is not recovered in the CCF analysis from both nights. We can constrain the abundance of HCN below $\sim$$10^{-4}$ from our data analysis and IRT tests. Furthermore, the IRTs indicate that both transits are not sensitive to CO, CO$_2$ , and CH$_4$, which explains their non-detection.

\subsection{KELT-17~b}

\begin{figure}[!hpbt]
    \centering
    \includegraphics[width=\linewidth]{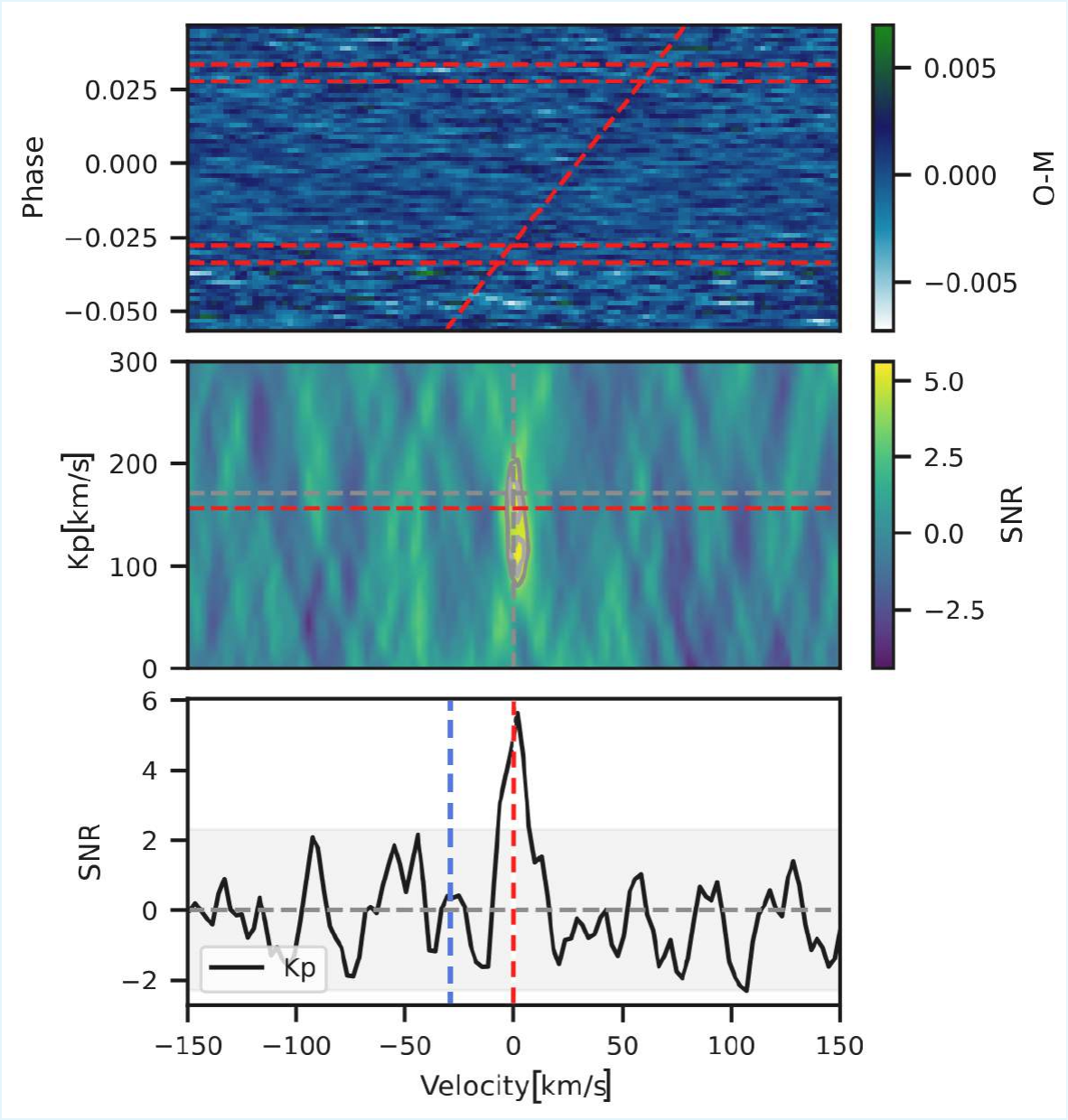}
    \caption{Result of the CC analysis of KELT-17~b on night of 26 January 2019   showing a detection of of FeH. The dashed red line of the K$_\mathrm{p} $ map marks the value of the K$_\mathrm{p}$ where we reach the maximum S/N. The interpretation of the dashed lines is as in Fig. 4.}
    \label{fig:FeHKELT2}
\end{figure}

For night 1, 26 orders were used and the sixth iteration of \sysrem provided the best results. We did not find any trace of the molecules studied ( H$_2$O, CO, CO$_2$, HCN, FeH, CH$_4$, TiO, VO, and OH).
On night 2, 31 orders were used and the best results were obtained after ten iterations of \sysrem. The Kp-map for FeH (Fig~\ref{fig:FeHKELT2}) shows a $\sim 5.3\sigma$ significant peak at $\rm K_p\,=\,145\,km/s$. No other significant signals are detected for the other molecules. As the first night presents useless data, the combination of both nights erases this detection.

The IRTs indicate that night 2 is significantly more sensitive to the presence of FeH than night 1, which is consistent with the CCF analyses. Therefore, we define the $5.3\sigma$ signal from the second night as a detection. The IRTs of the other molecules suggest that both transits do not have the quality to retrieve their signal with any significance.
Thus, the results from the CC analysis and IRTs for both nights and all molecules are consistent.

\begin{figure*}[!hpbt]
    \centering
    \includegraphics[width=\linewidth]{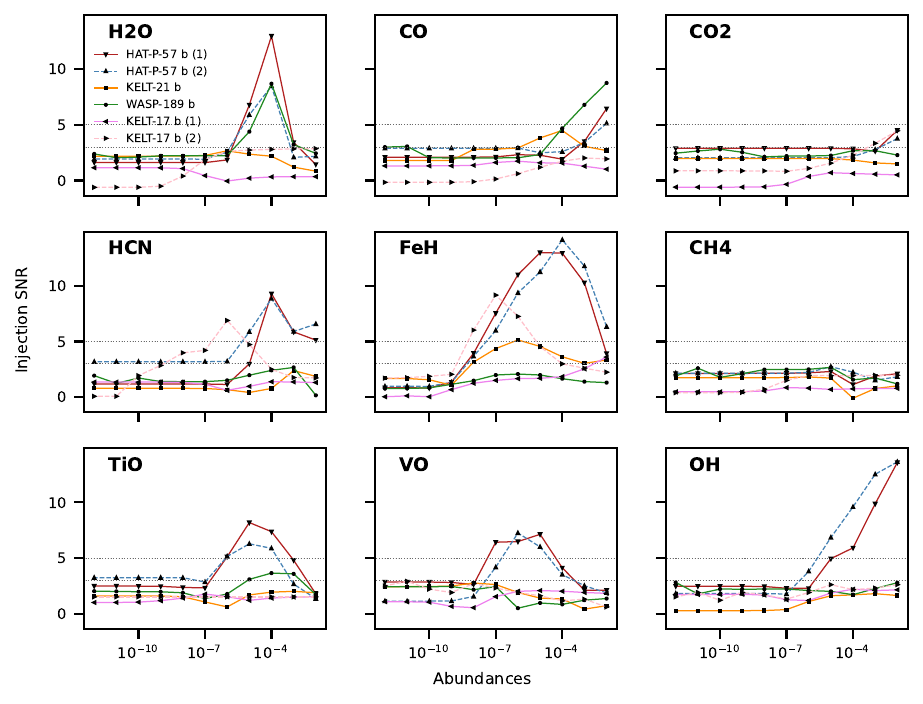}
    \caption{ Injection and recovery results for each inspected molecule (labelled in each panel) and each planet transit (colour coded for each individual transit according to the legend). We explored mass-fraction abundances (in molecular mean ratio (mmr)) from 10$^{-12}$ to 10$^{-2}$. The dotted horizontal lines indicate the 2.5 and 5 S/N values, respectively, in each panel.
    }
    \label{fig:context2}
\end{figure*}

\subsection{KELT-21~b}

For the single transit of KELT-21$~$b, we used 16 orders and the ninth iteration of \sysrem. We found no evidence of any molecular signature. The lack of features is supported by the IRTs. The injected H20, CO$_2$, CH$_4$, HCN, TiO, VO, OH signals are not recovered with significance ($<$3$\sigma$).
The IRTs also indicate a sensibility for CO of $10^{-5}$, which can be related to telluric contamination, and $10^{-8}$ for FeH, which is consistent with the behaviour of the molecule in high-temperature environments \citep{2018Parmentier}.

\subsection{WASP-189~b}

\begin{figure*}[!hpbt]
    \centering
    \includegraphics[width=\columnwidth]{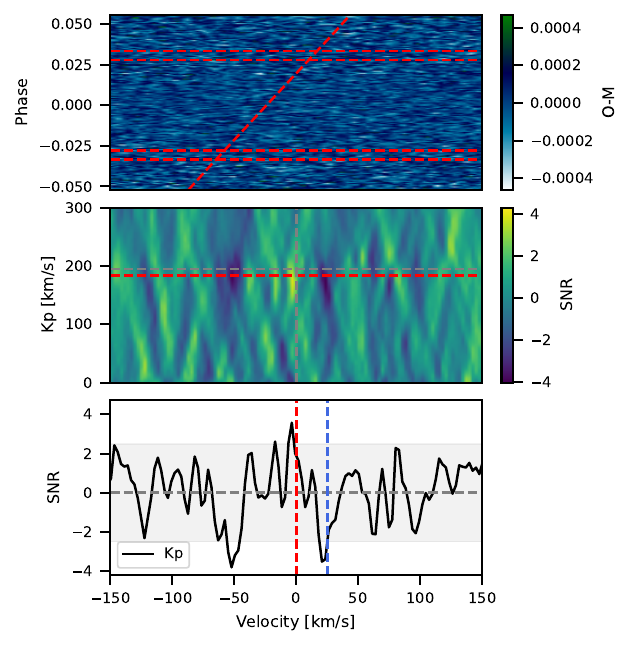}
    \includegraphics[width=\columnwidth]{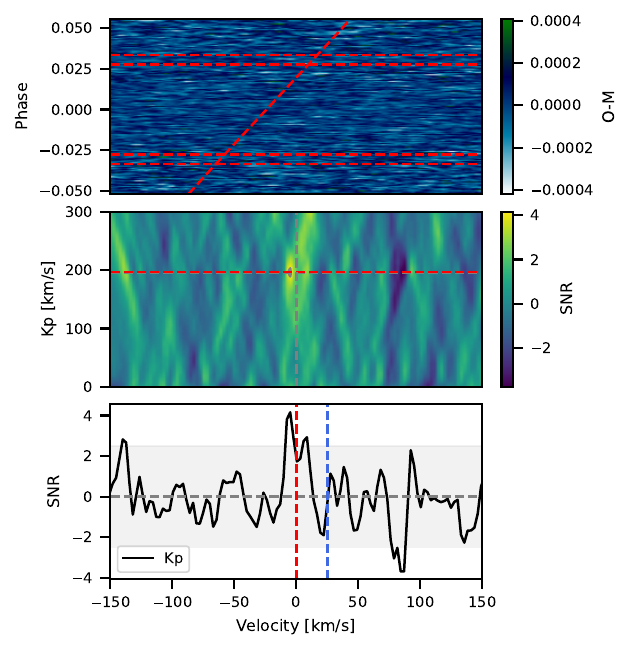}
    
    \caption{Result of the WASP-189$~$b observation CC analysis showing a tentative detection of H$_2$O on left panel and CO on right panel. The dashed red line of the K$_\mathrm{p} $ map marks the value of the K$_\mathrm{p}$ where we reach the maximum S/N. The interpretation of the dashed lines is as in Fig. 4.}
    \label{fig:wasp189}
\end{figure*}

For the single night of WASP-189~b, we used 24 orders and the ninth iteration of \sysrem. The CC analysis shows tentative molecular signals for H$_2$O (4.3$\sigma$) and CO (4.1$\sigma$), as can be seen in Figure \ref{fig:wasp189}. IRTs show that our data is sensitive enough for our tentative detection but does not show detectable spectral features for the rest of the molecules at any abundance, even in the case of atmospheres made up of more than 10$\%$ of the molecule.

\section{Discussion}

\subsection{HAT-P-57$~$b}
We only report a tentative $\rm H_2O$ detection from our molecular study. This detection is partially in agreement with previous studies on this planet. \cite{2020Kesseli} looked for FeH in 12 different exoplanets, including this planet, and reported no atmospheric detections. Additionally, \cite{Stangret2021} also reported no molecular or atomic absorptions in the optical wavelength range. However, detections have been made in similar UHJs to HAT-P-57$~$b: \cite{changeat2022five} found $\mathrm{ H_2O}$, FeH, and $\mathrm{CO_2}$ in HAT-P-7$~$b and \cite{2025Finnerty} report $\mathrm{ H_2O}$ and CO detections in MASCARA-2$~$b, both of them with a similar $\rm T_{eq}$ to our planet. Therefore, our tentative H$_2$O signal is in agreement with those findings, stressing the need to re-observe HAT-P-57$~$b and confirm our results. Further observations will help to confidently refute the presence of other molecules, such as FeH, CO, $\mathrm{CO_2}$, TiO, VO, or OH.

\subsection{KELT-17$~$b}

This work presents the first atmospheric study of KELT-17$~$b that analyses the NIR wavelength range. We report a 5.3$\sigma$ FeH signal, but no signatures for H$_2$O, CO$_2$, CO, CH$_4$, HCN, TiO, VO, and OH.
Previously, \cite{Stangret2021} analysed the optical transmission spectrum and found no absorption features for the same molecules inspected in this work. In particular, FeH has only been detected in a few other UHJs: WASP-103$~$b, KELT-9$~$b, and HAT-P-7$~$b \citep{changeat2022five}, which all have a similar temperature to our planet. Therefore, our result on KELT-17b adds a new unique data point in the FeH overview.

\subsection{KELT-21$~$b}

For the first time, we analyse the atmosphere of KELT-21$~$b in the NIR and find no atmospheric signal for FeH, H$_2$0, CO$_2$, CH$_4$ , and HCN in transmission. Our results are in agreement with the CC results from the optical transmission spectrum presented by \cite{Stangret2021}. They did not find evidence of molecular or atomic species in the optical wavelength range. Interestingly, \cite{Stangret2021} found an upper limit for the presence of $\mathrm{H_2O}$, while our IRT is unable to obtain a significant signal in both \giano-B datasets.

\subsection{WASP-189$~$b}
We found no traces of FeH, CO$_2$, CH$_4$, HCN, TiO, VO, or OH in the CC analysis, which is consistent with the results from the IRTs (Fig.\,\ref{fig:context2}). According to the ExoAtmospheres database, there is only one previous study for transmission spectroscopy in the NIR on this planet \citep{2025Vaulato}. Transmission spectroscopy studies in the optical range have been prolific in detecting metallic species, such as \citealp{2023Prinoth} where they detect TiO. Moreover, CO is detected in the emission spectrum of the planet \citep{yan2022detection, Lesjak2024, 2025sanchez} and H$_2$O is detected in emission \citep{2025sanchez}. These last two molecules are also detected (CO) and tentatively detected (H$_2$O) in our results, both showing doubled peaked signals. This might resemble the H$_2$O detection in WASP-127\,b  \citep{2025nortmann}, demonstrated to be related to wind jets or kinematics on the planet. The other molecular results are not surprising due to the detection of Fe \citep{2023Prinoth,2024Deibert,2025Vaulato}, which may indicate that FeH is dissociated. The FeH molecule has a strong bond, thus if it is dissociated, other molecules may not be present, apart from H$_2$O and CO \citep{2023Prinoth}.

\subsection{Injection and recovery tests}
We explored the molecular sub- to super-solar abundances with injection and recovery tests. The only molecule for which we are unable to set a detection threshold for all four planets is $\mathrm{CH_{4}}$. This may indicate that our observations are not sensitive to $\mathrm{CH_{4}}$.

In general, our analyses show that variations in elemental abundances can significantly affect the detectability of FeH. These variations suggest that such adjustments can bring the model closer to the observed transmission spectrum of the planet, as seen in Figure \ref{fig:context2}. Monte-Carlo-informed cross-correlation analyses, or high-resolution retrieval frameworks (which are out of the scope of this work), may offer a more robust pathway to infer the physical and chemical properties of these exoplanets beyond the confirmation of individual molecules and exploration of atmospheric winds. Nevertheless, these results are meant to show upper limits in our data sensitivity, so further and deeper studies may be needed to show abundance profiles.

\subsection{Molecules in the UHJ regime}

\begin{figure*}[!hpbt]
    \centering
    \includegraphics[width=\linewidth]{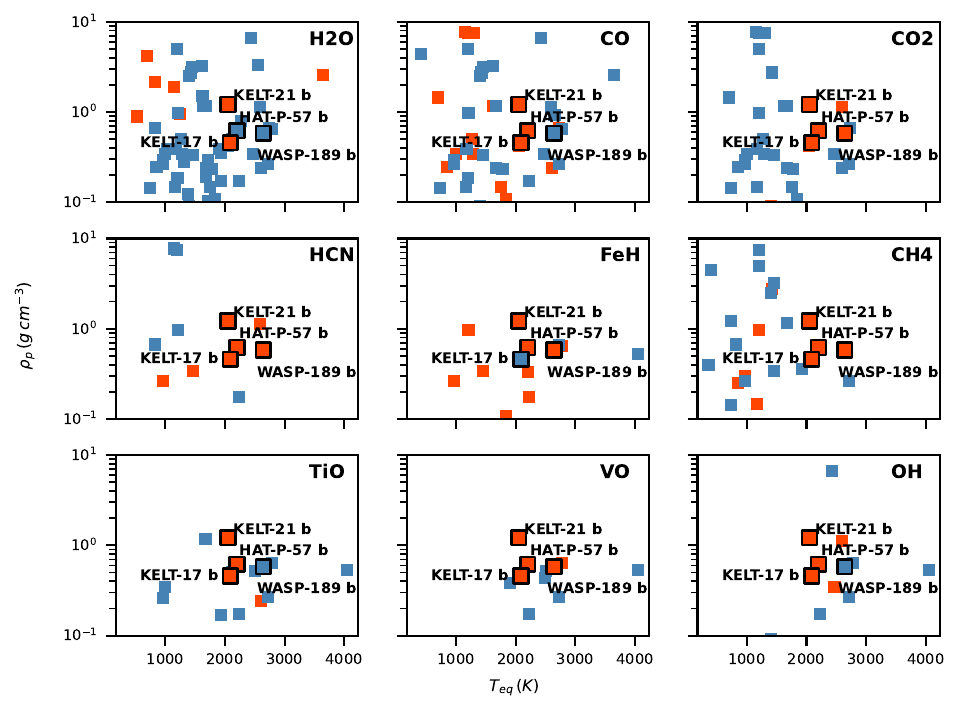}
    \caption{Planetary density (in logarithm scale) vs. equilibrium temperature ($T_{\rm eq}$ ) for the Jupiter-sized planets, with atmospheric studies of H$_2$O (top left), CO (top central), CO$_2$ (top right), CH$_4$ (middle left), FeH (middle central), HCN (middle right), TiO (bottom left), VO (bottom central), and OH (bottom right). We marked  the planets with detected signals of each molecule in green, while they are coloured in orange where no positive results were found. The planets studied in this work are highlighted and labelled in each panel.}
    \label{fig:context}
\end{figure*}

We only report the detection of FeH for KELT-17$~$b and the tentative detections of H$_2$O for HAT-P-57$~$b, and H$_2$O and CO  for WASP-189$~$b. All the other analyses resulted in non-detections. To put our findings in context, we compiled the atmospheric studies on Jupiter-sized planets (planets over 7 $\rm M_\oplus$ from the ExoAtmospheres database\footnote{Visited on 25 March 2026, \url{https://research.iac.es/proyecto/exoatmospheres/index.php}.}. Figure\,\ref{fig:context} presents the planet density versus equilibrium temperature diagram for each inspected molecule. The plotted planets are colour-coded according to the atmospheric results. We note that there is no keyword in ExoAtmospheres to classify the entries in transmission or emission studies.

Figure\,\ref{fig:context} does not show a clear trend as a function of planet density, with both detections and non-detections spanning the full range of density values. However, we do find some tentative trends related to the equilibrium temperature of the planets.
H$_2$O detections were found for almost all planets below $\sim$2750\,K. Therefore, this temperature limit trend is in agreement with the tentative detections for HAT-P-57$~$b and WASP-189$~$b. A similar trend was already pointed out by \cite{Stangret2022}. The non-detections for KELT-21\,b and KELT-17\,b can be explained as being due to the low quality of our observations (Fig. \ref{fig:context2}).

For CO, there are detections all over the diagram, even at the highest sampled temperatures ($>$3000\,K). This might be related to the covalent bond between the C and the O atoms.
The majority of the CO$_2$ detections come from planets with equilibrium temperatures $\leq$2000\,K, i.e. they are not considered UHJs. Only four low-density UHJs present CO$_2$ signals, surrounded by non-detections. A similar scenario is depicted by CH$_4$, where the detections are limited to cold and low-density planets.

HCN and FeH molecules show a scarcity of detections and explored planets as well. The few detections of HCN come for planets at equilibrium temperatures of $\sim$1000\,K and the only UHJ where HCN has been detected is HD$~$209458$~$b (\citealp{giacobbe2021five}, in transmission). On the other hand, the case of FeH is the most interesting as FeH signatures only appear in three planets beyond the barrier of UHJs ($\sim$2000\,K). It is expected that future studies of this molecule in UHJs will result in further detections. The coldest planet with a FeH detection is our target KELT-17\,b, which expands the low temperature regime where FeH can be detected.

TiO appears to be a widespread species among ultra-hot Jupiters (UHJs), with evidence of its presence reported in the atmospheres of a large fraction of the currently studied planets. In contrast, VO exhibits a more heterogeneous distribution, being detected only in a subset of UHJs. OH appears to show a stronger dependence on planetary temperature, with detections mainly associated with the hottest planets, typically with temperatures above $\rm \sim2500,K$.

\section{Conclusions}

In this work, we have explored the presence of molecules (H$_2$O, CO, CO$_2$, CH$_4$, FeH, HCN, TiO, VO, and OH) in the transmission spectra of four UHJ planets: HAT-P-57\,b, KELT-17\,b, KELT-21\,b, and WASP-189\,b. In particular, this is the first NIR transmission spectroscopy study for KELT-17b, KELT-21\,b, and WASP-189\,b. We also assessed the sensitivity of our data through injection and recovery tests and placed our results in the broader context of atmospheric detections in UHJs. The main results of this work are:

\begin{enumerate}
    \item We report a tentative detection of H$_2$O in HAT-P-57\,b at a significance of 4.3$\sigma$ in one of the two observed transits. The signal is not reproduced on the second night or in the combined dataset, probably because of differences in data quality. No other molecules are detected for this target on both nights.

    \item We detect FeH in KELT-17 b at 5.3$\sigma$, but it was only recovered on the higher-quality night. KELT-17\,b is the third UHJ and the coldest planet where FeH has been reported, which is a weird detection. No evidence of H$_2$O, CO, CO$_2$, CH$_4$, or HCN is found.

    \item No molecular species are detected in KELT-21 b. Injection--recovery tests indicate that the \giano-B data are not sensitive to mass-fraction abundances of 10$^{-12}$ to 10$^{-2}$.

    \item We made a tentative detection of H$_2$O (4.3$\sigma$) and CO (4.1$\sigma$) in WASP-189$~$b observations, which agree with the detections made in emission studies \citep{Lesjak2024, yan2022detection, 2025sanchez}.
    
    \item Injection--recovery tests show that molecular detectability strongly depends on both atmospheric abundance and data quality. For several molecules ($\rm H_2O,\, CO,\, CO_2,\, CH_4,\, FeH,\, HCN,\, TiO,\, VO,\, and\, OH$), the data are only sensitive to unrealistically high abundances, which may explain the absence of detections even for species commonly detected in similar planets.

    \item From an atmospheric perspective, our results support current expectations for UHJs: common molecules such as H$_2$O, CO, TiO, VO, and OH only appear when the data quality is sufficient; CO$_2$, CH$_4$ and HCN are rarely detected probably due to their dissociation and low abundance; and FeH emerges as a promising molecule in the hottest atmospheres, consistent with the limited number of previous detections. In general, we can see that, apart from CO and FeH, the other molecules show a presence limit around 3000 K, which is expected theoretically \citep{baxter2020transition}.

    \item The detection of FeH in KELT-17 b, and CO and the tentative detection of H$_2$O in WASP-189$~$b, with the tentative signals of H$_2$O in HAT-P-57 b, demonstrate the potential of NIR high-resolution transmission spectroscopy using ground-based 4-m-class telescopes for probing molecular chemistry in the most extreme gas giants. Further observations at a higher S/N and expanded wavelength coverage will be essential to confirm and extend the findings from this work.

\end{enumerate}

\begin{acknowledgements}
We acknowledge financial support from the Agencia Estatal de Investigaci\'on of the Ministerio de Ciencia e Innovaci\'on MCIN/AEI/10.13039/501100011033 and the ERDF “A way of making Europe” through project PID2021-125627OB-C32, and from the Centre of Excellence “Severo Ochoa” award to the Instituto de Astrofisica de Canarias.

This material is based upon work supported by the National Aeronautics and Space Administration under Grant Number  80NSSC20K0257 for the XRP program issued through the Science Mission Directorate.

We acknowledge the use of the ExoAtmospheres database during the preparation of this work.

This research has made use of the NASA Exoplanet Archive, which is operated by the California Institute of Technology, under contract with the National Aeronautics and Space Administration under the Exoplanet Exploration Program.
This research has made use of NASA’s Astrophysics Data System.

\end{acknowledgements}

\bibliographystyle{aa} 
\bibliography{citations}

\begin{appendix}
\label{Sec:Appendix}
\onecolumn
\section{Helium triplet analyses of $\rm KELT-17\,b$ and $\rm KELT-21\,b$}
\label{App: He analyses}

\begin{figure*}[h]
    \centering
    \includegraphics[width=1\linewidth]{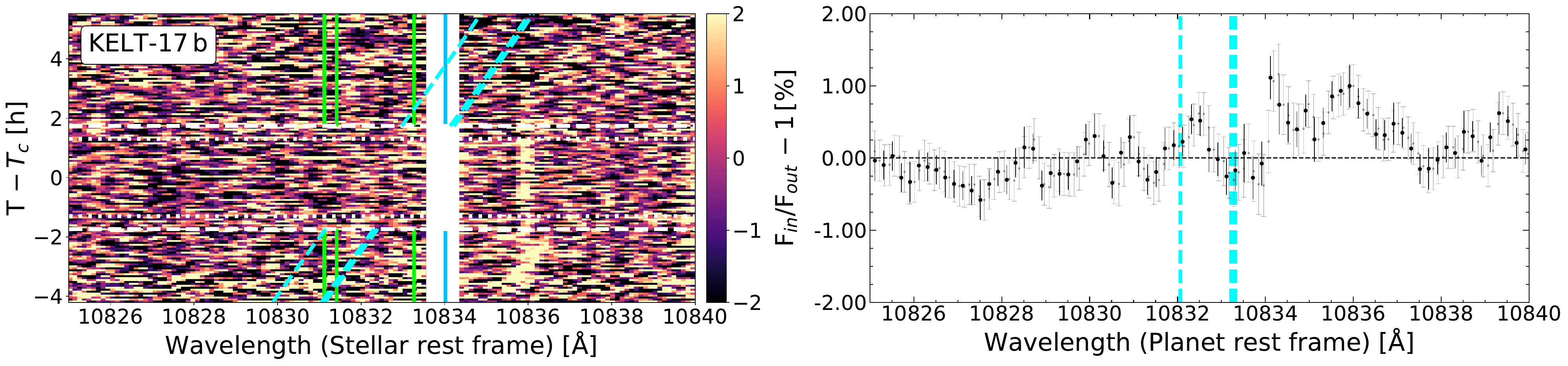}
    \includegraphics[width=1\linewidth]{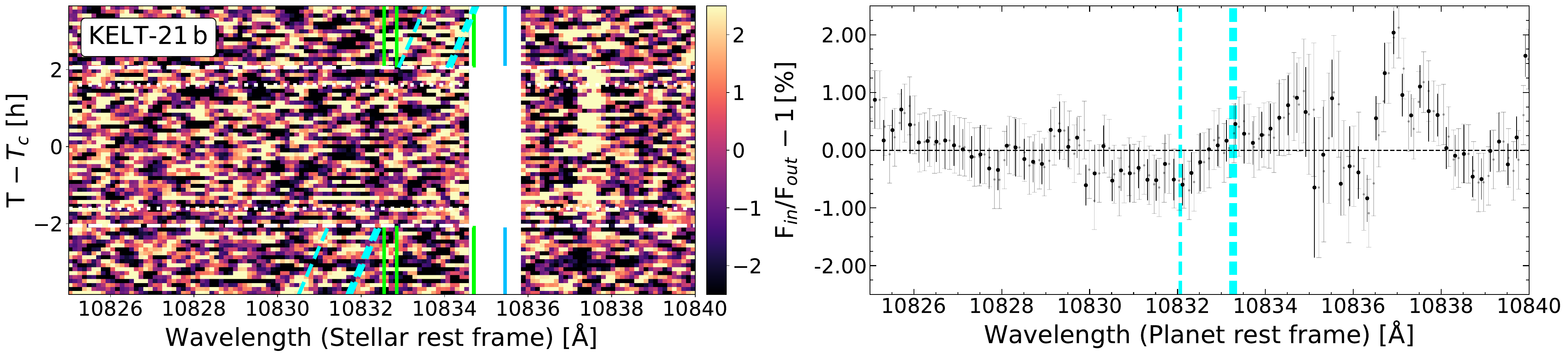}
    \caption{\label{Fig: Helium Figs}
    Residual maps and transmission spectra around the \ion{He}{I} NIR triplet lines for KELT-17\,b (top panels) and KELT-21\,b (bottom panels) observations.
    Left panels: Residual maps in the stellar rest frame. The time since mid-transit time ($T_{\rm c}$) is shown on the vertical axis, and the relative absorption is colour-coded. The dashed and dotted white horizontal lines indicate the different contacts during the transit. The dashed cyan tilted lines indicate the predicted trace of the planetary signals. The solid green vertical lines indicate the position of the OH emission telluric lines. The solid blue line indicates the position of the H$_2$O absorption telluric line. The residuals of the telluric lines are masked.
    Right panels: Planet transmission spectra in the planet rest frame. We show the original data in light grey and the data binned by 0.2\,$\AA$ in black. The dashed cyan vertical lines indicate the \ion{He}{I} triplet lines positions. All the wavelengths in this figure are referenced in a vacuum.
    }
\end{figure*}

We performed single-line transmission spectroscopy technique to study the possible absorption from the Helium near-infrared triplet line at 1083\,nm. The HAT-P-57\,b and WASP-189\,b Helium analyses have already been discussed in \cite{2024Mopys}. Here, we aim to contribute to the population study of this line by reporting, for the first time, results on KELT-17\,b and KELT-21\,b. In summary, we computed a high SNR stellar spectrum by combining all the out-of-transit spectra. Then, we divided the spectroscopic time series by the high SNR stellar spectrum to obtain the residual map (left panels in Figure\,\ref{Fig: Helium Figs}). Finally, we combined the residual spectra taken fully in transit to compute the transmission spectra (right panels in Figure\,\ref{Fig: Helium Figs}). We accounted for the different rest frames during the analysis. For further details, see the methodology applied to the \giano-B data in \cite{2024Mopys}.

The residual maps from both planets show no absorption signal aligned with the planetary trace (Figure\,\ref{Fig: Helium Figs}). Both transmission spectra are dominated by noise, with no evidence of planetary absorption.
We derive a 3$\sigma$ upper limit to the He triplet absorption of 0.7\,\% for KELT-17\,b, and of 1\,\% for KELT-21\,b.

\end{appendix}

\end{document}